\documentclass[sigconf, nonacm]{acmart}

\usepackage{mathtools}
\usepackage{amsmath}
\allowdisplaybreaks[4]
\usepackage[shortlabels]{enumitem}

\usepackage{etoolbox}
\AtEndPreamble{
  \theoremstyle{acmdefinition}
  \newtheorem{assumption}[theorem]{Assumption}
  \newtheorem{remark}[theorem]{Remark}
}

\newcommand{\N}{\mathbb{N}}
\newcommand{\R}{\mathbb{R}}
\newcommand{\X}{\mathcal{X}}
\newcommand{\U}{\mathcal{U}}
\newcommand{\W}{\mathcal{W}}

\newcommand{\dint}[2]{[#1;#2]}

\newcommand{\defeq}{\coloneqq}
\newcommand{\interior}[1]{#1^{\circ}}
\newcommand{\cl}[1]{\overline{#1}}

\newcommand{\leb}{\lambda}

\newcommand{\Post}{\operatorname{Post}}
\newcommand{\Prob}{\mathbb{P}}
\newcommand{\indic}{\mathbf{1}}

\newcommand{\UnivPre}{\mathbb{A}\mathrm{pre}}

\newcommand{\LPre}{\mathbb{L}\mathrm{pre}}
\newcommand{\ALPre}{\mathbb{AL}\mathrm{pre}}
\newcommand{\Win}{\mathbb{W}\mathrm{in}}
\newcommand{\Fair}{\mathrm{Fair}}

\newcommand{\Until}{\mathbin{\textsf{U}}}

\newcommand{\RA}{\edit{\varphi_{Q \Until T}}} %

\usepackage{tikz}
\usetikzlibrary{arrows.meta}

\definecolor{axisgray}{HTML}{555555}
\definecolor{setgreen}{HTML}{8FC997}
\definecolor{setblue}{HTML}{3A70A8}
\definecolor{trajred}{HTML}{C0503F}

\newcommand{\SES}[1]{\textcolor{black}{#1}}
\colorlet{editcolor}{black}
\newcommand{\edit}[1]{{\color{editcolor}#1}} %
\begin{document}

\title[Topological Fairness in Continuous-State Systems]{Topological Fairness in Continuous-State Systems}

\author{Mahdi Nazeri}
\affiliation{%
  \institution{University of Oxford \& MPI-SWS}
  \city{Oxford}
  \country{United Kingdom}
}
\email{mahdi.nazeri@cs.ox.ac.uk}

\author{Alessandro Abate}
\affiliation{%
  \institution{University of Oxford}
  \city{Oxford}
  \country{United Kingdom}
}
\email{alessandro.abate@cs.ox.ac.uk}

\author{Anne-Kathrin Schmuck}
\affiliation{%
  \institution{MPI-SWS}
  \city{Kaiserslautern}
  \country{Germany}
}
\email{akschmuck@mpi-sws.org}

\author{Sadegh Soudjani}
\affiliation{%
  \institution{University of Birmingham \& MPI-SWS}
  \city{Birmingham}
  \country{United Kingdom}
}
\email{s.soudjani@bham.ac.uk}

\begin{abstract}
Since the early days of probabilistic programs, fairness has been used to abstract randomness in software. Intuitively, a fair scheduler reproduces the almost-sure behavior of a random choice based solely on the set of possible outcomes, independent of their (often unknown) distribution. As this abstraction attaches fairness to individual transitions, it becomes ineffective over uncountable state spaces, where every single successor has probability zero. 

We overcome this limitation by imposing fairness on open sets instead. This leads to the novel notion of \emph{topological fairness}: every basic open set---with respect to a countable basis of the state-space topology---that lies in the set of possible successors infinitely often must be visited infinitely often. On discrete state spaces where singletons form a topological basis, this recovers classical fairness. 

We show that topological fairness carries the intuition from program abstraction to continuous state spaces. For systems with additive stochastic disturbances, almost every trajectory is topologically fair under every control policy, based solely on the support of the disturbance. Reasoning over all topologically fair trajectories is therefore a sound proof principle for almost-sure properties of the original system. Unlike in the finite-state case, however, it is not complete: topological fairness cannot capture almost-sure behavior that hinges on the relative likelihoods of disturbance outcomes.

Finally, we show that topological fairness supports synthesis directly on the continuous state space. For reach-avoid objectives, we give a nested fixed point over subsets of the state space which, for a broad class of systems, computes the fair winning domain \emph{exactly} together with a memoryless winning strategy.
Our case studies show that this winning domain is substantially larger than the one obtained with currently available techniques\edit{, which detour through a finite abstraction and certify a winning domain that depends on the discretization in a non-monotone way}.
\end{abstract}

\maketitle

\section{Introduction}
\label{sec:intro}

In controller synthesis for \emph{continuous-state} dynamical systems, stochastic disturbances are typically handled in one of two ways. On one hand, a \emph{probabilistic model} can be used which represents the likelihood of each disturbance
realization. Its analysis is precise, but may require numerical transition
probabilities and operations over probability distributions. On the other hand, an \emph{adversarial
model} retains only the set of possible disturbance values, i.e., the support
of each probabilistic choice, and is hence independent of its often unknown distribution. The adversarial model is often much simpler to analyze, but places no restriction on how disturbances occur over time. Hence, synthesis under an adversarial model may reject a controller because of disturbance realizations that form a
probability-zero event under the stochastic model.

For \emph{finite-state} systems, a middle ground between these two extremes is well known. A \emph{fair model} is an adversarial model in which a transition that is enabled infinitely often is also taken infinitely often. 
This fairness condition holds almost surely under every probabilistic model
with the same transition support~\cite{HartSharirPnueli1983,Pnueli1983,BaierKwiatkowska1998,BaierKatoen2008}.
Moreover, for $\omega$-regular objectives, the fair and almost-sure winning
domains coincide. Almost-sure synthesis can therefore be reduced to solving a
two-player game under strong transition fairness, for which efficient symbolic
algorithms exist~\cite{BanerjeeEtAl2022,BanerjeeEtAl2023,MajumdarEtAl2023CAV}.

For \emph{continuous-state} systems, this middle ground does not yet exist. Current fairness assumptions need a finite set of events to attach to, which a continuous state space does not offer. When the disturbance admits a density, there are uncountably many possible successor states, each of which with probability zero. Requiring every one of them to occur infinitely often is impossible, while requiring only that successors stay in the support of the disturbance imposes no restriction.
Existing fairness-based controller synthesis techniques for continuous-state
systems circumvent this problem through abstraction.
Here, the system is first replaced by a finite abstraction, whose abstract states supply the missing events, and a fair game is then solved on the abstraction \cite{MajumdarMallikSoudjani2020,MajumdarMallikSchmuckSoudjani2021,MajumdarMallikSchmuckSoudjani2024}.
This construction can introduce conservatism that depends sensitively on the
chosen discretization.
This paper therefore asks the following question:

\begin{center}
\emph{Does a fair model for stochastic continuous-state systems exist?}
\end{center}

We answer this question positively by introducing \emph{topological fairness}, a notion that carries the benefits of finite-state fair models over to continuous-state systems.

\paragraph{Contribution.}
We define topological fairness with respect to a countable basis of the
Euclidean topology, consisting of open balls with rational centers and radii.
A basis element is enabled at a state--input pair if it is contained in the
corresponding set of possible successors, and a play is fair if every basis
element that is enabled infinitely often is visited infinitely often.
On a discrete state space, where singletons form a topological basis, the
definition recovers classical strong fairness.

Our first main result establishes that topological fairness is indeed a
fair model in the sense above. For systems with additive stochastic
disturbances, trajectories are topologically fair almost surely, under every
control policy and under every disturbance distribution over the same
support.
Topological fairness hence excludes only a probability-zero set of
disturbance behaviors, and the controller does not need to enforce it as an
additional assumption. The proof relies on the fact that each fixed basis
element has a uniform positive probability of being visited whenever it is
enabled, together with a conditional form of the Borel--Cantelli lemma.

As a consequence, topological fairness induces a game semantics that lies
between the conventional adversarial and probabilistic interpretations. For
every Borel-measurable objective over infinite trajectories, including
reachability, safety, and objectives induced by $\omega$-regular
specifications, we establish the hierarchy
\[
    \Win_{0}^{\mathrm{adv}}
    \subseteq
    \Win_{0}^{\mathrm{f}}
    \subseteq
    \Win_{0}^{\mathrm{p}},
\]
where the three sets are, respectively, the adversarial, fair, and almost-sure
winning domains. Importantly, \emph{both} inclusions can be strict. The first
strictness is the reason why fair models are less conservative than
adversarial ones: states that are losing against a fully adversarial
disturbance become winning once the disturbance is restricted to fair
behaviors. The second strictness marks a difference from the finite-state
setting, where the fair and the almost-sure winning domains coincide
\cite{BanerjeeEtAl2023}. There, almost-sure behavior is fully determined by
the support of the probabilistic choices. On continuous state spaces, this is
no longer true: almost-sure behavior may hinge on the relative likelihoods of
disturbance outcomes, e.g., on a drift induced by the expected disturbance,
which no fairness condition formulated over the support alone can capture.
The fair model is hence a sound, but not complete, qualitative surrogate for
the probabilistic one.

\SES{Our second main result exploits topological fairness for controller
synthesis without constructing a finite abstraction. For reach-avoid
objectives, we introduce a nested fixed point whose predecessor operator
combines invariance with progress under fair disturbances. A parameter
$\epsilon\geq0$ controls the required progress: for $\epsilon>0$, the
fixed point gives a sound under-approximation,
while at $\epsilon=0$, under appropriate regularity and
finite-convergence conditions, it exactly characterizes the fair winning
domain and yields a memoryless winning strategy.}

\SES{We illustrate the fixed-point construction on one analytically
solvable example and two numerical case studies. In all these systems,
the fair winning domain is substantially larger than the adversarial
winning domain, which consists only of the target. A comparison with a state-of-the-art fairness-based abstraction method shows that the winning domain certified through abstraction depends non-monotonically on the grid resolution: refining the grid can shrink the certified domain to the empty set, while further refinement recovers it. This behavior reflects a sensitivity of the current state-of-the-art methods to grid alignment that our abstraction-free construction avoids.}

\paragraph{Related work.}
Fairness has a long history in the verification of concurrent and reactive
systems; see \cite{Francez1986,Kwiatkowska1989,vanGlabbeekHoefner2020} for
monograph and survey treatments. Classical notions include unconditional,
weak, and strong fairness, which differ in when an action or transition is
required to occur \cite{QueilleSifakis1983}. The fair model of this paper
is inspired by strong transition fairness, which requires an enabled transition to be
taken infinitely often whenever it is enabled infinitely often
\cite{AminofBallKupferman2005}.

The abstraction of model randomness by fairness goes back to work on probabilistic
programs and processes. Pnueli's extreme fairness \cite{Pnueli1983} and the
framework of Baier and Kwiatkowska \cite{BaierKwiatkowska1998} establish that
satisfaction along all fair executions is a sound proof principle for
almost-sure satisfaction of linear-time properties. Completeness is more
delicate even for finite systems. In nondeterministic planning, Aminof et
al.~\cite{AminofDeGiacomoRubin2020} distinguish \emph{state-action fairness},
which constrains the traces admitted by the environment, from
\emph{stochastic fairness}, which excludes arbitrary probability-zero sets of
traces; the two agree for reachability but can differ for temporally extended
goals. The second strictness in our hierarchy exposes a related gap on
continuous state spaces, already for reach-avoid objectives.

On finite graphs, the fair model is also algorithmically effective.
Qualitative synthesis for $\omega$-regular objectives in finite stochastic
games is well developed
\cite{ChatterjeeDeAlfaroHenzinger2005,ChatterjeeHenzinger2012}, and the fact
that qualitative behavior depends only on the transition support underlies
reductions in which probabilistic choices are replaced by fair adversarial
choices. A direct symbolic algorithm for stochastic Rabin games implements
this reduction through an almost-sure predecessor operator
\cite{BanerjeeEtAl2022}, and strong transition fairness has since led to
direct symbolic algorithms for $\omega$-regular games \cite{BanerjeeEtAl2023},
Zielonka-style algorithms for odd-fair parity games \cite{SaglamSchmuck2023},
and decremental algorithms for fair B\"uchi games
\cite{SaglamSchmuckTsyrempilon2025}. %
In all of these algorithms fairness is attached to individual edges of a finite graph, while topological fairness is defined over the continuous successor set.

\looseness=-1
For continuous-state stochastic systems, formal synthesis commonly relies on
finite abstractions, interval-valued Markov models, or probabilistic games
\cite{DBLP:journals/automatica/AbatePLS08,DBLP:journals/siamads/SoudjaniA13,SvorenovaEtAl2017,LaurentiEtAl2020,DutreixHuhCoogan2022,LavaeiEtAl2022}.
The detour discussed above combines such abstractions with fair games:
probabilistic transitions are replaced by fair adversarial choices on a
finite game that uses only the support of the stochastic kernel
\cite{MajumdarMallikSoudjani2020,MajumdarMallikSchmuckSoudjani2021,
MajumdarMallikSchmuckSoudjani2024}, and a toolchain combines solvers for fair
and stochastic Rabin games with symbolic abstractions
\cite{MajumdarEtAl2023CAV}.
The present work provides a direct continuous-state semantic foundation for this construction.
It gives fairness a semantics directly on the uncountable state
space and computes the reach-avoid winning domain without partitioning the state space or bounding transition probabilities between partitions.

Two further lines of work relate fairness to topology or to continuous
dynamics, but differently from ours. V\"olzer and Varacca
\cite{VolzerVaracca2012} use Baire category on the space of executions to
characterize fairness as a topologically large set of runs; we instead use
the topology of the state space to decide which successor neighborhoods are
subject to fairness, so topological fairness is a pathwise generalization of
transition fairness rather than a measure of the size of a set of executions.
Zhou et al.~\cite{ZhouEtAl2025} impose user-specified fairness constraints on
epistemic disturbance actions of uncertain dynamical systems and solve the
resulting LTL synthesis problem on a finite adversarial Markov decision
process obtained by partitioning state, input, and disturbance spaces. Our
fairness obligations are instead induced by the topology of the continuous
successor sets, hold almost surely for the stochastic disturbance, and are
defined before any abstraction. The two approaches thus address complementary
roles of fairness: prescribed regularity of epistemic uncertainty versus
probability-consistent behavior of stochastic disturbances.

\section{Preliminaries}
\label{sec:prelims}
We denote by $\N$ the natural numbers (including zero) and by $\R$ the real numbers. For $a \leq b$ in $\N$, let $\dint{a}{b} \defeq \{n \in \N \mid a \leq n \leq b\}$. For a set $A$, $A^{*}$ and $A^{\omega}$ denote the sets of finite and infinite sequences over $A$. For a subset $S \subseteq \R^{n}$, $\interior{S}$ denotes its topological interior, i.e., the largest open subset of $\R^{n}$ contained in $S$, and $\cl{S}$ denotes its closure. Finally, $\leb$ denotes the $n$-dimensional Lebesgue measure.

This paper focuses on stochastic dynamical systems with additive disturbances, as defined next.
\begin{definition}[Stochastic Systems]
    \label{def:system}
    A \emph{stochastic dynamical system} is a tuple $\mathcal{S} = (\X, \U, (\W, \mathfrak{B}(\W), \mu), f)$, where:
    \begin{itemize}[leftmargin=1.5em]
        \item $\X \subseteq \R^{n}$ is a Borel set representing the continuous \emph{state space};
        \item $\U$ is a finite \emph{input space};
        \item $(\W, \mathfrak{B}(\W), \mu)$ is a \emph{probability space} over the disturbance, where $\W \subset \R^{n}$ is non-empty and compact, $\mathfrak{B}(\W)$ is its Borel $\sigma$-algebra, and $\mu$ is a probability measure with positive density almost everywhere on $\W$; and
        \item $f \colon \X \times \U \to \R^{n}$ is a Borel measurable \emph{transition map} satisfying
            \begin{equation*}
                f(x,u) + \W \subseteq \X \qquad \text{for every } (x,u) \in \X \times \U.
            \end{equation*}
    \end{itemize}
    The system evolves according to
    \begin{equation}
        \label{eq:dynamics}
        x(k+1) = f\bigl(x(k), u(k)\bigr) + w(k), \qquad k \in \N,
    \end{equation}
    where $x(k) \in \X$, $u(k) \in \U$, and $(w(k))_{k \in \N}$ is an i.i.d.\ sequence of random variables drawn according to $\mu$. Given $x \in \X$ and $u \in \U$, the set of possible successor states is
    \begin{equation}
        \label{eq:post}
        \Post(x,u) \defeq f(x,u) + \W.
    \end{equation}
\end{definition}

We can interpret the evolution of $\mathcal{S}$ as an interaction in which the controller first chooses $u \in \U$ and the environment (disturbance) then chooses the successor state from $\Post(x,u)$. \edit{Which successors the environment may select over time is what distinguishes the \emph{adversarial} model, where every selection is allowed, from the \emph{probabilistic} model where the selection must respect $\mu$. The \emph{fair} model of Section~\ref{sec:topological-fairness} abstracts the latter into a fair choice. The game
graph defined next formalizes this interaction. It records only the successor
sets $\Post(x,u)$, i.e., the support of the disturbance, and thus serves as the
common structure for all environment models.}

\begin{definition}[Induced Game Graph]
    \label{def:game-graph}
    The \edit{stochastic dynamical} system $\mathcal{S}$ induces the infinite bipartite two-player game graph $G(\mathcal{S}) = (V_{0}, V_{1}, E)$, with Player~$0$ \edit{(the controller)} vertices $V_{0} = \X$ and Player~$1$ \edit{(the environment)} vertices $V_{1} = \X \times \U$. The edge relation $E = E_{0} \cup E_{1}$ is defined by
    \begin{align}
        E_{0} &\defeq \{\, (x, (x,u)) \in V_{0} \times V_{1} \mid u \in \U \,\}, \label{eq:edges-0} \\
        E_{1} &\defeq \{\, ((x,u), x') \in V_{1} \times V_{0} \mid x' \in \Post(x,u) \,\}. \label{eq:edges-1}
    \end{align}
\end{definition}

Let $V \defeq V_{0} \cup V_{1}$, and let $V^{*}$ and $V^{\omega}$ denote the set of finite and infinite trajectories, respectively.
A \emph{Player~$0$ strategy} is a measurable map
$\rho_{0} \colon V^{*} \!\cdot V_{0} \to \U$,
and a \emph{Player~$1$ strategy} is a map $\rho_{1} \colon V^{*} \!\cdot V_{1} \to V_{0}$ satisfying $\rho_{1}(h(x,u)) \in \Post(x,u)$. A strategy is \emph{memoryless} if it depends only on the current vertex, reducing to $\rho_{0} \colon V_{0} \to \U$ and $\rho_{1} \colon V_{1} \to V_{0}$ with $\rho_{1}(x,u) \in \Post(x,u)$. For $x_{0} \in V_{0}$, strategies $\rho_{0}, \rho_{1}$ induce a unique play
\begin{equation*}
    \pi_{x_{0}}(\rho_{0}, \rho_{1}) = x_{0}(x_{0}, u_{0})x_{1}(x_{1}, u_{1})x_{2}\dots \in V^{\omega},
\end{equation*}
where $u_{i} = \rho_{0}\bigl(x_{0}(x_0,u_0)\dots x_{i}\bigr)$ and $x_{i+1} = \rho_{1}\bigl(x_{0}(x_0,u_0)\dots (x_{i}, u_{i})\bigr)$, or simply $u_{i} = \rho_{0}(x_{i})$ and $x_{i+1} = \rho_{1}(x_{i}, u_{i})$ when memoryless. A \emph{winning condition} is a set $\varphi \subseteq V^{\omega}$, and Player~$0$ wins a play $\pi$ if and only if $\pi \in \varphi$.
\section{Topological Fairness}
\label{sec:topological-fairness}
In this section, we present our core contribution: the formulation of
fairness over continuous state spaces. \edit{Section~\ref{sec:fair-plays} defines topologically fair plays, shows that the definition recovers strong fairness on discrete state spaces, and proves the main result: plays of a stochastic dynamical system are almost surely fair. Section~\ref{subsec:hierarchy} then shows that the resulting fair winning domain lies between the adversarial and the
almost-sure winning domains, and both inclusions can be strict.}

\subsection{Topologically Fair Plays}
\label{sec:fair-plays}
The key idea behind topological fairness is to focus on the basis of the
state-space topology instead of unaccountably many states in the continuous space. This forms a countable family of events since the Euclidean state space $\R^{n}$ is second-countable and hence admits a countable topological basis. We fix a canonical enumeration
\begin{equation}
    \label{eq:countable-basis}
    \mathcal{B} = (B_{j})_{j \geq 1}
\end{equation}
of open Euclidean balls with rational centers and positive rational radii\edit{; every open subset of $\R^{n}$ is a union of elements of $\mathcal{B}$}. \edit{Topological fairness can now be defined relative to this basis.}

\begin{definition}[Topologically Fair Play]
    \label{def:fair-play}
    \color{editcolor}
    Let $G(\mathcal{S})$ be the game graph induced by a stochastic dynamical system $\mathcal{S}$, and let $\pi = x_{0}(x_{0}, u_{0})x_{1}(x_{1}, u_{1})\dots$ be a play over $G(\mathcal{S})$. Then $\pi$ is called \emph{(topologically) fair} if, for every $j \geq 1$,
    \begin{equation}
        \label{eq:basis-fairness}
        |E_{j}(\pi)| = \infty \implies |S_{j}(\pi)| = \infty,
    \end{equation}
    where
    \begin{subequations}\label{eq:basis-notions}
        \begin{align}
            E_{j}(\pi) &\defeq \{\, i \in \N \mid B_{j} \subseteq \Post(x_{i}, u_{i}) \,\} \quad\text{and} \label{eq:basis-enabled}\\
            S_{j}(\pi) &\defeq \{\, i \in E_{j}(\pi) \mid x_{i+1} \in B_{j} \,\} \label{eq:basis-served}
        \end{align}
    \end{subequations}
    are the sets of steps at which the basis element $B_{j}$ is \emph{enabled} and \emph{visited}, respectively. 
\end{definition}

That is, every basis element that is enabled infinitely often is visited infinitely often. On finite state spaces, this is exactly strong state fairness.

\begin{remark}[Strong State Fairness]
    \label{rem:strong-fairness}
    On a finite state space with the discrete topology, the singletons $\{v\}$ form a countable topological basis. For this basis, the enabling condition $\{v\} \subseteq \Post(x_{i}, u_{i})$ reduces to $v \in \Post(x_{i}, u_{i})$ and the visiting condition $x_{i+1} \in \{v\}$ to $x_{i+1} = v$, so the fairness condition~\eqref{eq:basis-fairness} enforces that every state that is enabled infinitely often is visited infinitely often. This is strong state fairness. On a continuous state space, the analogous pointwise condition is incompatible with absolutely continuous disturbances: singletons have probability zero and are almost surely never
visited even when they are enabled infinitely often. Basic open sets are the events that keep the condition meaningful.
\end{remark}

As the main result of this section, we now show that Definition~\ref{def:fair-play} has the defining property of a fair model: almost every play of a stochastic dynamical system is fair, under every control strategy. To state this, recall that under Definition~\ref{def:system}, any measurable controller strategy $\rho_{0}$ and any initial state $x_{0} \in \X$ induce a probability measure $\Prob_{x_{0}}^{\rho_{0}}$ over the set of plays $V^{\omega}$, equipped with the product $\sigma$-algebra.

\begin{theorem}[Fairness of Stochastic Systems]
    \label{thm:prob-fairness}
    For any initial state $x_{0} \in \X$ and any (possibly history-dependent or randomized) Player~$0$ strategy $\rho_{0}$, a play $\pi = x_{0}(x_{0}, u_{0})x_{1}(x_{1}, u_{1})\dots$ is fair with probability~$1$, i.e.,
        $\Prob_{x_{0}}^{\rho_{0}}\bigl( \{\, \pi \in V^{\omega} \mid \pi \text{ is fair} \,\} \bigr) = 1$.
\end{theorem}

We must show that fairness holds for every basis ball $B_j$: if it is enabled infinitely often, it is visited infinitely often as well. We first show that as a ball is repeatedly enabled, the probability of its being visited accumulates over time. The following lemma establishes this by proving that whenever $B_j$ is enabled, the probability of visiting it in the next step is bounded below by a constant $p_j > 0$ independent of $(x,u)$. We then use L\'evy's conditional Borel--Cantelli lemma (see~\cite[Section~12.15]{williams1991probability}) to prove that as $B_j$ is enabled infinitely often and the sum of these probabilities diverges, $B_j$ is visited infinitely often almost surely. This is true for each basis ball and thus concludes the proof. 

\begin{lemma}[Bounded visiting probability]
    \label{lem:pj}
    For every $j \geq 1$, there exists $p_j > 0$ such that for all $(x,u) \in \X \times \U$ with $B_j \subseteq \Post(x,u)$,
    \[
        \mu\bigl(\{\, w \in \W \mid f(x,u) + w \in B_j \,\}\bigr) \geq p_j.
    \]
\end{lemma}
\begin{proof}
    Let $B_j$ have center $c_j$ and radius $r_j > 0$. Partition $\R^n$ into cubes of diameter $r_j/2$. Since $\W$ is bounded, only finitely many cubes $Q_1, \dots, Q_m$ lie in $\W$, and each has $\mu(Q_i) > 0$ because $\mu$ has positive density almost everywhere on $\W$. Set $p_j \defeq \min_{i \in \dint{1}{m}} \mu(Q_i) > 0$. Whenever $B_j \subseteq f(x,u) + \W$, the set $\{\, w \in \W \mid f(x,u) + w \in B_j \,\}$ is the ball $B(c, r_j) \subseteq \W$ centered at $c \defeq c_j - f(x,u)$. The unique cube $Q$ containing $c$ has diameter $r_j/2 < r_j$, so $Q \subseteq B(c, r_j) \subseteq \W$ giving $\mu(B(c, r_j)) \geq \mu(Q) \geq p_j$.
\end{proof}

\begin{proof}[\edit{Proof of Theorem~\ref{thm:prob-fairness}}]
    By Definition~\ref{def:fair-play}, a play $\pi$ is fair if and only if it belongs to the event
    \begin{equation*}
        \Fair \defeq \bigcap_{j=1}^{\infty} \Bigl( \{\, \pi \mid |E_{j}(\pi)| < \infty \,\} \cup \{\, \pi \mid |S_{j}(\pi)| = \infty \,\} \Bigr).
    \end{equation*}
    Since a countable intersection of events of probability~$1$ has probability~$1$, it suffices to show that for every fixed $j \geq 1$,
    \begin{equation}
        \label{eq:single-basis-fairness}
        \Prob_{x_{0}}^{\rho_{0}}\bigl( |E_{j}(\pi)| = \infty \text{ and } |S_{j}(\pi)| < \infty \bigr) = 0.
    \end{equation}

    Fix $j \geq 1$. At each step $k \in \N$, let $h_k \defeq x_0(x_0,u_0)\dots(x_k,u_k)$ and $\mathcal{F}_k \defeq \sigma(h_k)$, and define the $\mathcal{F}_{k+1}$-measurable event $A_k \defeq \{k \in S_j(\pi)\} = \{k \in E_j(\pi)\} \cap \{x_{k+1} \in B_j\}$.
    Because $w(k)$ is independent of $h_k$ and $B_j \subseteq \Post(x_k,u_k)$ whenever $k \in E_j(\pi)$, Lemma~\ref{lem:pj} gives
    \begin{equation*}
        \Prob\bigl( x_{k+1} \in B_{j} \bigm| h_{k} \bigr)
        = \mu\bigl(\{\, w \in \W \mid f(x_k, u_k) + w \in B_j \,\}\bigr)
        \geq p_{j} > 0
    \end{equation*}
    whenever $k \in E_j(\pi)$. Since $\{k \in E_j(\pi)\}$ is determined by $h_k$, it follows that
    $\Prob( A_k \mid h_{k} ) \geq p_{j} \cdot \indic_{\{k \in E_j(\pi)\}}$, and summing over $k$ gives
    \begin{equation*}
        \sum_{k=0}^{\infty} \Prob\bigl( A_k \bigm| h_{k} \bigr) \geq p_{j} \cdot |E_{j}(\pi)|.
    \end{equation*}
    Consequently, on $\{|E_{j}(\pi)| = \infty\}$ the sum of conditional probabilities diverges, and L\'evy's conditional Borel--Cantelli lemma implies that $A_k$ occurs for infinitely many $k$ almost surely, i.e., $|S_j(\pi)| = \infty$. This proves Eq.~\eqref{eq:single-basis-fairness}. The (countable) intersection over all balls gives $\Prob_{x_0}^{\rho_0}(\Fair) = 1$.
\end{proof}

The proof above only used the support of the disturbance and the fact that it is almost everywhere positive on it. Therefore, the result holds for every disturbance density with the same support. We showed that topological fairness excludes only a probability-zero set of disturbance behaviors, and reasoning over all fair plays is a sound proof principle for almost-sure properties of the stochastic system, similarly to the finite-state case.

\subsection{Hierarchy of Winning Domains}
\label{subsec:hierarchy}

\edit{Each environment model of Section~\ref{sec:prelims} induces a corresponding winning domain: the set of states from which the controller has a strategy that satisfies a winning condition $\varphi$
according to
the semantics of that model.
We define the three winning domains side by side and then show that they form a hierarchy.}
\begin{definition}[Winning Domains]
    \label{def:fair-winning}
    \color{editcolor}
    Let $G = G(\mathcal{S})$ be the game graph induced by a stochastic dynamical system $\mathcal{S}$, let $\varphi \subseteq V^{\omega}$ be a winning condition, and write $\pi \defeq \pi_{x_{0}}(\rho_{0}, \rho_{1})$. The \emph{adversarial}, \emph{fair}, and \emph{almost-sure winning domains} are
    \begin{align*}
        \Win_{0}^{\mathrm{adv}}(G,\varphi) &\defeq \bigl\{\, x_{0} \in V_{0} \mid \exists \rho_{0} \; \forall \rho_{1} : \pi \in \varphi \,\bigr\}, \\
        \Win_{0}^{\mathrm{f}}(G,\varphi) &\defeq \bigl\{\, x_{0} \in V_{0} \mid \exists \rho_{0} \; \forall \rho_{1} : \pi \text{ is fair} \implies \pi \in \varphi \,\bigr\}, \\
        \Win_{0}^{\mathrm{p}}(G,\varphi) &\defeq \bigl\{\, x_{0} \in V_{0} \mid \exists \rho_{0} : \Prob_{x_{0}}^{\rho_{0}}(\varphi) = 1 \,\bigr\}.
    \end{align*}
\end{definition}
\edit{In the adversarial winning domain, the controller must win against every Player~$1$ strategy, i.e., every selection of (environment) successors from $\Post(x,u)$. In the fair winning domain, unfair plays are disregarded, so the controller only has to win against fair behaviors of the environment. In the almost-sure winning domain, the successors are drawn according to $\mu$, and the controller must win with probability one. As this implies that the task for the system player (Player 0) gets easier as we move from an adversarial, through a fair, to a probabilistic environment model, it follows that the three winning domains are nested.}

\begin{theorem}[Hierarchy of Winning Domains]
    \label{thm:hierarchy}
     For every game graph $G = G(\mathcal{S})$ induced by a \edit{stochastic dynamical} system $\mathcal{S}$ and every Borel winning condition $\varphi \subseteq V^{\omega}$,
    \begin{equation}
        \label{eq:hierarchy}
        \Win_{0}^{\mathrm{adv}}(G,\varphi) \subseteq \Win_{0}^{\mathrm{f}}(G,\varphi) \subseteq \Win_{0}^{\mathrm{p}}(G,\varphi).
    \end{equation}
\end{theorem}

\edit{For finite-state systems, the same hierarchy is well known, with a notable difference: the left inclusion is strict in general, whereas the right one is an equality for $\omega$-regular winning conditions~\cite{BanerjeeEtAl2023}, since almost-sure behavior is fully determined by the support of the probabilistic choices. Theorem~\ref{thm:hierarchy} shows that both inclusions carry over to continuous state spaces, and their proofs follow the finite-state case: the left inclusion holds because the controller has to win against fewer plays, and the right one follows from Theorem~\ref{thm:prob-fairness}, since fair plays have probability one. The difference lies in strictness. The left inclusion is strict for the same reason as in the finite-state case (Example~\ref{ex:adv-vs-fair}): fairness forces the environment to eventually visit an enabled target region, which an adversary could avoid forever. The right inclusion, however, can also be strict on continuous state spaces (Example~\ref{ex:fair-vs-prob}). Here, almost-sure behavior may hinge on the relative likelihoods of disturbance outcomes, e.g., on a drift induced by the expected disturbance. %
The fair model is hence less conservative than the adversarial one, and a sound, but not complete, qualitative surrogate for the probabilistic one.}

\begin{proof}[Proof of Theorem~\ref{thm:hierarchy}]
    \noindent\textbf{First inclusion ($\Win_{0}^{\mathrm{adv}} \subseteq \Win_{0}^{\mathrm{f}}$).}\quad
    Let $x_{0} \in \Win_{0}^{\mathrm{adv}}(G,\varphi)$, witnessed by a Player~$0$ strategy $\rho_{0}$ such that
    \begin{equation*}
        \pi_{x_{0}}(\rho_{0}, \rho_{1}) \in \varphi \qquad \text{for every Player~$1$ strategy } \rho_{1}.
    \end{equation*}
    Fix an arbitrary $\rho_{1}$ and let $\pi = \pi_{x_{0}}(\rho_{0}, \rho_{1})$. Since $\pi \in \varphi$ unconditionally, the implication $(\pi \text{ is fair} \implies \pi \in \varphi)$
    holds trivially, regardless of whether $\pi$ is fair. As $\rho_{1}$ was arbitrary, the same $\rho_{0}$ witnesses $x_{0} \in \Win_{0}^{\mathrm{f}}(G,\varphi)$ by Definition~\ref{def:fair-winning}.

    \medskip
    \noindent\textbf{Second inclusion ($\Win_{0}^{\mathrm{f}} \subseteq \Win_{0}^{\mathrm{p}}$).}\quad
    Let $x_0 \in \Win_{0}^{\mathrm{f}}(G,\varphi)$ witnessed by a Player~$0$ strategy $\rho_0$.
    Under the stochastic dynamics, running $\rho_0$ from $x_0$ produces a fair play almost surely (Theorem~\ref{thm:prob-fairness}).
    Every play produced by the dynamics under $\rho_0$ is also a valid play of the game (realized by a Player~$1$ strategy $\rho_1$ that replays its transitions), and $\rho_0$ wins on every fair play.
    Therefore $\Prob_{x_0}^{\rho_0}(\varphi) = 1$, so $x_0 \in \Win_{0}^{\mathrm{p}}(G,\varphi)$.
\end{proof}
The following two examples (illustrated in Figure~\ref{fig:examples}) show that both inclusions can be strict. Both use the reach-avoid (safe-reachability) winning condition $\RA$ with safe set $Q$ and target set $T$, formally defined in Section~\ref{sec:synthesis}.

\begin{figure}[t]
\centering
\begin{tikzpicture}[
    scale=0.72,
    axis/.style={
        axisgray,
        line width=0.6pt,
        -{Stealth[length=5pt,width=4pt]}
    },
    tick/.style={
        axisgray,
        line width=0.45pt
    },
    dot/.style={
        circle,
        fill=trajred,
        inner sep=0pt,
        minimum size=5.5pt
    },
    lab/.style={
        font=\normalsize
    },
    subcap/.style={
    font=\small,
    align=center
},
]

  \begin{scope}[x=4cm, y=1cm]

    \draw[axisgray, line width=0.7pt]
  (-1,1.3) -- (-1,1.6) -- (1,1.6) -- (1,1.3);
\node[lab, above=1pt] at (0,1.6)
  {$\Post(x,u)=[-1,\,1]$};
  
    \fill[setgreen] (0,-0.22) rectangle (0.5,0.22);
    \fill[setblue]  (0.125,-0.22) rectangle (0.375,0.22);
    \draw[tick] (0,-0.25) -- (0,0.25);
    \draw[tick] (0.5,-0.25) -- (0.5,0.25);

    \draw[axis] (-1.27,0) -- (1.27,0)
      node[below=7pt, font=\normalsize] {$x$};

    \node[lab, below=6pt] at (-1,0)   {$-1$};
    \node[lab, below=6pt] at (0,0)    {$0$};
    \node[lab, below=6pt] at (0.25,0) {$T$};
    \node[lab, below=6pt] at (0.5,0)  {$0.5$};
    \node[lab, below=6pt] at (1,0)    {$1$};

    \node[lab] at (0.25,0.75) {$B$};

    \node[dot] at (-1,0) {};
    \node[dot] at (1,0) {};
    \node[lab, trajred] at (-0.87,0.75) {$x_k=-1$};
    \node[lab] at (1,0.70) {$x_0$};

    \node[subcap] at (0,-1.35) {
  \textbf{(a)}\
  $\Win_0^{\mathrm{adv}}\subsetneq\Win_0^{\mathrm{f}}$:
  An unfair play avoids $T$.
};
  \end{scope}

  \begin{scope}[
      shift={(-1.94cm,-3.55cm)},
      x=2.96cm,
      y=1cm
  ]

    \fill[setgreen] (0.5,-0.22) rectangle (2,0.22);
    \draw[tick] (0.5,-0.25) -- (0.5,0.25);
    \draw[tick] (2,-0.25) -- (2,0.25);
    \node[lab] at (1.25,0.75) {$T$};

    \draw[axis] (-1.08,0) -- (2.37,0)
      node[below=7pt, font=\normalsize] {$x$};

    \node[lab, below=6pt] at (0,0)   {$0$};
    \node[lab, below=6pt] at (0.5,0) {$0.5$};
    \node[lab, below=6pt] at (2,0)   {$2$};

    \draw[
        trajred,
        line width=1.5pt,
        -{Stealth[length=6pt,width=6pt]}
    ]
      (-0.58,0) -- (-0.97,0);

    \foreach \k in {0,...,5}
      \node[dot] at ({-0.1*\k},0) {};

    \node[lab, trajred] at (-0.55,0.75)
      {$x_k=-0.1k$};
    \node[lab] at (0.025,0.70) {$x_0$};

    \node[subcap] at (0.655,-1.35) {
  \textbf{(b)}\
  $\Win_0^{\mathrm{f}}\subsetneq\Win_0^{\mathrm{p}}$:
  A fair play avoids $T$.
};
  \end{scope}

\end{tikzpicture}
\caption{The plays of Examples~\ref{ex:adv-vs-fair} (a)
and~\ref{ex:fair-vs-prob} (b); the target $T$ is green. In (a), the
play stays at $x_k=-1$ and avoids $T$, but is unfair: the ball
$B\subseteq T$ is enabled at every step and never visited. In (b), the
play $x_k=-0.1k$ avoids $T$ and is fair: its successor sets drift away,
so no ball stays enabled.}
\label{fig:examples}
\end{figure}

\begin{example}[$\Win^{\mathrm{adv}}_0 \subsetneq \Win^{\mathrm{f}}_0$]
    \label{ex:adv-vs-fair}
    \color{editcolor}
    Figure~\ref{fig:examples}(a) shows a system on $\X = \R$ without control choice, $\U = \{u\}$, with $f \equiv 0$ and $\mu$ uniform on $\W = [-1,1]$. Every successor set is therefore $\Post(x,u) = [-1,1]$, independently of the current state. The objective is to reach $T = [0, \tfrac12]$ from $x_0 = 1$, with $Q = \R$.
    An adversarial environment can keep the play at $x_k = -1$ forever. This play never reaches $T$, so $x_0 \notin \Win_0^{\mathrm{adv}}(G, \RA)$.
    The play is not fair, however. The ball $B$ with center $\tfrac14$ and radius $\tfrac18$ lies inside $T$ and inside every successor set, so it is enabled at every step but never visited. In a fair play, $B$ is visited infinitely often, and in particular $T$ is reached. Hence $x_0 \in \Win_0^{\mathrm{f}}(G, \RA)$.
\end{example}

\begin{example}[$\Win^{\mathrm{f}}_0 \subsetneq \Win^{\mathrm{p}}_0$]
    \label{ex:fair-vs-prob}
    \color{editcolor}
    Figure~\ref{fig:examples}(b) shows the random walk $x(k+1) = x(k) + w(k)$ on $\X = \R$, again without control choice, with $\mu$ uniform on $\W = [-0.1, 0.5]$. The objective is to reach $T = [0.5, 2]$ from $x_0 = 0$, with $Q = \R$.
    The expected step $\mathbb{E}[w] = 0.2$ is positive, so $x(k) \to +\infty$ almost surely by the strong law of large numbers, and since no step exceeds $0.5$, the play cannot jump over $T$. Thus $T$ is reached almost surely, and $x_0 \in \Win_{0}^{\mathrm{p}}(G, \RA)$.
    Now consider the play that always takes the smallest step, $x_k = -0.1k$. It drifts to $-\infty$ and never reaches $T$. It is nevertheless fair: its successor sets $[x_k - 0.1,\, x_k + 0.5]$ drift away as well, so every basis ball is enabled only finitely often, and the fairness condition~\eqref{eq:basis-fairness} holds vacuously. Hence $x_0 \notin \Win_{0}^{\mathrm{f}}(G, \RA)$.
    Fairness only constrains which regions are visited while they keep being enabled; it cannot enforce the drift induced by the expected disturbance.
\end{example}
\section{Synthesis under Topological Fairness}
\label{sec:synthesis}
So far, we have introduced the topological notion of fairness for the continuous domain and established that it naturally holds almost surely for stochastic dynamical systems. In this section, we present our second main contribution which is a fixed-point algorithm to synthesize controllers under such a fairness assumption. We focus on the \emph{safe-reachability problem} (a.k.a. \emph{reach-avoid problem}): given a safe set $Q \subseteq V_{0}$ and a target set $T \subseteq Q$, Player~$0$ has to drive the system into the target set $T$ in finite time while ensuring the state remains within the safe set $Q$ at all intermediate steps until $T$ is reached. Formally, this objective corresponds to the reach-avoid winning condition
\begin{equation*}
    \RA \defeq \left\{\, \pi = x_{0}(x_{0},u_{0})x_{1}(x_{1},u_{1})\dots \;\middle|\;
    \begin{aligned}
        &\exists k \in \N : x_{k} \in T, \text{ and} \\
        &\forall i < k : x_{i} \in Q
    \end{aligned}
    \,\right\}.
\end{equation*}
\edit{This is a Borel winning condition, so Theorem~\ref{thm:hierarchy} applies, and Examples~\ref{ex:adv-vs-fair} and~\ref{ex:fair-vs-prob} show that both inclusions of the hierarchy can be strict already for reach-avoid winning conditions. The reach-avoid synthesis problem then asks for (i) a characterization of the fair winning domain $\Win_0^{\mathrm f}(G(\mathcal S), \RA)$, and (ii) a Player~$0$ strategy that wins against every fair play from every state in this domain.}

\subsection{Continuous Predecessor Operators}
\label{sec:prelims:cpre}
\edit{The fixed point of Section~\ref{sec:nested-fixpoint} is built from a predecessor operator that combines two requirements on a state--input pair: \emph{invariance}, i.e., all successors stay in a given set, and \emph{progress}, i.e., 
the successors overlap with a given set, thereby providing an opportunity for
progress. The soundness results below show that, under the stated assumptions,
fairness turns such repeated opportunities into eventual progress.
We now define these operators.}  
Let $S, \edit{P} \subseteq V_{0}$. We define operators on the induced game graph, for $\epsilon \geq 0$:
\begin{align*}
    \UnivPre(S)     &\defeq \{\, (x,u) \in V_{1} \mid \Post(x,u) \subseteq S \,\}, \\
    \LPre(S)        &\defeq \{\, (x,u) \in V_{1} \mid \Post(x,u) \cap \interior{S} \neq \emptyset \,\}, \\
    \ALPre(S,\edit{P})     &\defeq \{\, x \in V_{0} \mid \exists u \in \U : (x,u) \in \UnivPre(S) \cap \LPre(\edit{P}) \,\}, \\
    \LPre^{\epsilon}(S)    &\defeq \{\, (x,u) \in V_{1} \mid \leb\bigl(\Post(x,u) \cap \interior{S}\bigr) > \epsilon \,\}, \\
    \ALPre^{\epsilon}(S,\edit{P}) &\defeq \{\, x \in V_{0} \mid \exists u \in \U : (x,u) \in \UnivPre(S) \cap \LPre^{\epsilon}(\edit{P}) \,\},
\end{align*}
i.e.,
\begin{equation}
    \label{eq:alpre-eps-expanded}
    \ALPre^{\epsilon}(S,\edit{P}) = \left\{\, x \in V_{0} \;\middle|\;
    \begin{aligned}
        &\exists u \in \U : \Post(x,u) \subseteq S, \\
        &\leb\bigl(\Post(x,u) \cap \interior{\edit{P}}\bigr) > \epsilon
    \end{aligned}
    \,\right\}.
\end{equation}
\edit{Here, $\UnivPre$ captures invariance, $\LPre$ captures progress, and $\ALPre$ combines both with an existential choice of input. The parameter $\epsilon \geq 0$ quantifies the required overlap between the successor set and the progress set; for $\epsilon = 0$, a non-empty overlap with the interior suffices.} All defined operators are monotone in the set arguments in which they occur.
\edit{For the exact characterization of the fair winning domain, we require the dynamics and the reach-avoid sets to be definable in an o-minimal structure, which covers in particular semialgebraic sets and maps. We recall the basic notions and refer to}
\cite{vanDenDries1998} for a comprehensive discussion.
\begin{definition}[O-Minimal Structure $\R_{\mathrm{an},\exp}$]
    \label{def:ominimal}
    An \emph{o-minimal structure} $\mathcal{M} = (\mathcal{M}_{n})_{n \geq 1}$ over the ordered real field $(\R, +, \cdot, <)$ is a sequence where each $\mathcal{M}_{n}$ is a Boolean algebra of subsets of $\R^{n}$ (called \emph{definable sets}) such that:
    \begin{enumerate}[(1)]
        \item $\mathcal{M}_{n}$ contains all sets $\{\, x \in \R^{n} \mid P(x) = 0 \,\}$ and $\{\, x \in \R^{n} \mid P(x) > 0 \,\}$ for every polynomial $P \in \R[x_{1},\dots,x_{n}]$;
        \item $\mathcal{M}$ is closed under Cartesian products (if $A \in \mathcal{M}_{n}$ and $B \in \mathcal{M}_{m}$, then $A \times B \in \mathcal{M}_{n+m}$) and coordinate projections (if $A \in \mathcal{M}_{n+1}$, then $\pi(A) \in \mathcal{M}_{n}$ where $\pi(x_{1},\dots,x_{n},x_{n+1}) = (x_{1},\dots,x_{n})$);
        \item every definable subset in $\mathcal{M}_{1}$ is a finite union of points and open intervals.
    \end{enumerate}

    A function $g \colon A \subseteq \R^{n} \to \R^{m}$ is definable in $\mathcal{M}$ if its graph $\operatorname{graph}(g) = \{\, (x, g(x)) \mid x \in A \,\}$ belongs to $\mathcal{M}_{n+m}$.

    Throughout this work, we fix $\mathcal{M}$ to be the rich o-minimal structure $\R_{\mathrm{an},\exp}$, which is the expansion of $(\R, +, \cdot, <)$ by all restricted real-analytic functions and the unrestricted exponential function. The structure $\R_{\mathrm{an},\exp}$ contains all polynomials, rational functions, exponentials, logarithms, power functions, restricted trigonometric functions, and analytic functions on compact domains and its definable sets are Borel. We will use two essential properties that come with this structure:
    \textbf{(P1)} a finite union of definable sets with empty interior has empty interior;
    \textbf{(P2)} for every definable $g \colon A \to \R^m$ there is a finite partition of $A$ into definable sets on each of which $g$ is continuous.
\end{definition}
For the remainder of this section, we make the following regularity assumption.
\begin{assumption}[System Regularity]
    \label{ass:geometry}
    Let $\mathcal{M}$ be the o-minimal structure of Definition~\ref{def:ominimal}. We assume that:
    \begin{enumerate}[(i)]
        \item The safe set $Q \subseteq \X$ and the target set $T \subseteq Q$ are bounded and definable in $\mathcal{M}$.
        \item For each control input $u \in \U$, the transition map $x \mapsto f(x,u)$ is definable in $\mathcal{M}$.
        \item The disturbance set is a hyperrectangle $\W = [a,b] \defeq \{\, w \in \R^{n} \mid a_i \leq w_i \leq b_i \text{ for all } i \in \dint{1}{n} \,\}$ for vectors $a, b \in \R^n$ with $a_i < b_i$ for all $i$.
    \end{enumerate}
\end{assumption}

Under Assumption~\ref{ass:geometry}(iii), every $\Post(x,u) = f(x,u) + \W$ is a closed non-empty hyperrectangle. Since $\interior{S}$ is open, any non-empty intersection $\Post(x,u) \cap \interior{S}$ contains an open ball and thus has positive measure. Therefore, $\LPre^0 = \LPre$ and $\ALPre^0 = \ALPre$.

\subsection{Fixed-Point Algorithm}
\label{sec:nested-fixpoint}
\edit{We now introduce the nested fixed point over a game $G(\mathcal S)$ for a reach-avoid winning condition $\RA$ as} 
\begin{equation}
    \label{eq:main-fixpoint}
    Z_\epsilon \defeq \nu Y.\, \mu X.\, \Bigl( T \cup \bigl( Q \cap \ALPre^{\epsilon}(Y, X) \bigr) \Bigr)
\end{equation}
where $X,Y$ are subsets of the continuous state space.

\SES{Its shape mirrors the fixed point for finite-state reach-avoid games under strong transition fairness~\cite{BanerjeeEtAl2023}, with the edge-based predecessor operators replaced by the set-based operators of Section~\ref{sec:prelims:cpre}. For $\epsilon>0$, boundedness of $Q$, rectangularity of $\W$, finite convergence, and Borel measurability of the induced rank-decreasing strategy suffice to ensure that $Z_\epsilon$ is contained in the fair winning domain; in particular, this result does not require o-minimal definability. At $\epsilon=0$, under Assumptions~\ref{ass:geometry} and~\ref{ass:convergence}, $Z_0$ coincides with the fair winning domain.}

\edit{Intuitively, the inner least fixed point of \eqref{eq:main-fixpoint} collects the states in $Q$ from which some input keeps all successors inside the candidate set $Y$ and makes progress towards states already collected; under fairness, this progress is eventually realized. The outer greatest fixed point shrinks $Y$ until it consists only of states from which the inner fixed point is reachable without leaving $Y$.} For a candidate set $Y \subseteq \X$, let
\begin{equation*}
    \Phi_\epsilon(Y) \defeq \mu X.\, \Bigl( T \cup \bigl( Q \cap \ALPre^{\epsilon}(Y, X) \bigr) \Bigr),
\end{equation*}
which is monotone in $Y$, hence $Z_\epsilon = \nu Y.\, \Phi_\epsilon(Y)$. The inner least fixed point is computed via the increasing sequence of approximants
\begin{equation}
    \label{eq:inner-approximants}
    X_{m+1}(Y) \defeq T \cup \bigl( Q \cap \ALPre^{\epsilon}(Y, X_{m}(Y)) \bigr), \quad m \in \N,
\end{equation}
initialized with $X_{0}(Y) \defeq \emptyset$.
\edit{Dually}, the outer greatest fixed point is computed by the approximants
\begin{equation}
    \label{eq:outer-iterates-eps}
    Y_{k+1} \defeq \Phi_\epsilon(Y_{k}), \qquad k \in \N,
\end{equation}
initialized with $Y_{0} \defeq \X$.
By monotonicity of $\Phi_\epsilon$, $Y_0 \supseteq Y_1 \supseteq \cdots$, and $T \subseteq Y_k$ for all $k$.
\edit{On a finite game graph, both sequences stabilize after at most as many steps as there are vertices. On a continuous state space, there is no such bound: the inner sequence is increasing and the outer one decreasing, but neither needs to reach its limit in finitely many steps. In the remainder of this section we therefore assume that the fixed-point iteration over \eqref{eq:main-fixpoint} terminates, and verify it in each case study presented in Section~\ref{sec:examples}.}

\begin{assumption}[Finite Convergence of Fixed Points]
    \label{ass:convergence}
    \edit{For the value of $\epsilon$ under consideration, all fixed-point iterations terminate: for every $k \in \N$ there is $K_k \geq 1$ with $X_{K_k+1}(Y_k) = X_{K_k}(Y_k)$, s.t.\ $Y_{k+1} = X_{K_k}(Y_k)$, and there is $N \geq 1$ with $Y_{N+1} = Y_N$, s.t.\ $Z_\epsilon = Y_N$.}
\end{assumption}
\SES{For $Y=Z_\epsilon$, we write $X_m \defeq X_m(Z_\epsilon)$ and $K \defeq K_N$.
Then $Z_\epsilon=X_K$ and $X_1=T$.}
\edit{As in the finite-state case, the inner approximants induce a rank on $Z_\epsilon$ and a memoryless strategy that, outside $T$, keeps the play in $Z_\epsilon$ and makes progress towards lower rank.}
\begin{definition}[Rank and rank-decreasing strategy]
    \label{def:rank-strategy}
    \color{editcolor}
    Under Assumption~\ref{ass:convergence}, the \emph{rank} of $x \in Z_\epsilon$ is $r(x) \defeq \min \{\, m \in \dint{1}{K} \mid x \in X_{m} \,\}$. The memoryless Player~$0$ strategy $\rho_{0}^{Z}$ selects, for $x \in Z_\epsilon \setminus T$, the first input $u$ in a fixed enumeration of $\U$ with
    \begin{equation}
        \label{eq:rank-decreasing-input}
        \Post(x,u) \subseteq Z_\epsilon \qquad \text{and} \qquad \leb\bigl( \Post(x,u) \cap \interior{X_{r(x)-1}} \bigr) > \epsilon,
    \end{equation}
    and an arbitrary input (e.g., first input $u$ in a fixed enumeration of $\U$) for $x \in T$.
\end{definition}
The strategy is well-defined: we have $r(x) = 1$ iff $x \in T$, and for $x \in Z_\epsilon \setminus T$ we have $r(x) \geq 2$ and $x \in \ALPre^{\epsilon}(Z_\epsilon, X_{r(x)-1})$ by the definition of $X_{r(x)}$, so an input satisfying~\eqref{eq:rank-decreasing-input} exists.
Recall that Player~$0$ strategies are required to be Borel measurable\edit{, i.e., the sets $\{x \in Z_\epsilon \setminus T \mid \rho_0^Z(x) = u\}$, $u \in \U$ must be Borel. For $\epsilon = 0$ this directly follows (Lemma~\ref{lem:struct}) and for $\epsilon > 0$ it is added as an assumption in Theorem~\ref{thm:sound-eps}.} For $x \in Z_\epsilon$, we use the shorthand notation $F(x) \defeq f(x,\rho_0^Z(x))$.

\subsection{Soundness}
\label{sec:soundness}
We now prove the soundness of \eqref{eq:main-fixpoint}, i.e., $Z_\epsilon\subseteq \Win_0^{\mathrm f}(G(\mathcal S), \RA)$. To this end, we fix $\rho_0^Z$ as the rank-decreasing strategy of Definition~\ref{def:rank-strategy} and show that every fair play consistent with $\rho_0^Z$ from a state in $Z_\epsilon$ reaches $T$ within $Q$. We first show this result for $\epsilon > 0$, where the positive intersection required by the progress condition gives a direct argument, and then for $\epsilon = 0$, where we use o-minimality to prove soundness. Both parts share the safety argument.
\begin{lemma}[Safety]
    \label{lem:safety}
    Let $\pi$ be a play from $x_0 \in Z_\epsilon$ \edit{consistent with} $\rho_0^Z$, and suppose $x_0,\dots,x_i \notin T$. Then $x_0,\dots,x_i \in Z_\epsilon \setminus T \subseteq Q$, $x_{i+1} \in Z_\epsilon$, and each $u_j = \rho_0^Z(x_j)$, $j \leq i$, satisfies~\eqref{eq:rank-decreasing-input}. Consequently, $\pi \in \RA$ if and only if $\pi$ visits $T$.
\end{lemma}
\begin{proof}
    We have $x_0 \in Z_\epsilon$, and if $x_j \in Z_\epsilon \setminus T$, then $u_j$ satisfies~\eqref{eq:rank-decreasing-input}, so $x_{j+1} \in \Post(x_j,u_j) \subseteq Z_\epsilon$. Induction gives the first claims, except for $Z_\epsilon \setminus T \subseteq Q$, which holds since $Z_\epsilon = X_K = T \cup (Q \cap \ALPre^\epsilon(Z_\epsilon, X_{K-1})) \subseteq T \cup Q = Q$. If $\pi$ first visits $T$ at step $k$, then $x_0,\dots,x_{k-1} \in Q$, so $\pi \in \RA$. The converse is trivial.
\end{proof}

\subsubsection{Soundness of \eqref{eq:main-fixpoint} for $\epsilon > 0$}
\label{sec:soundness-eps}

\begin{theorem}[Soundness for $\epsilon > 0$]
    \label{thm:sound-eps}
    \SES{
    Fix $\epsilon > 0$. Suppose that $Q$ and $T$ are Borel, $Q$ is
    bounded, Assumption~\ref{ass:geometry}(iii) and
    Assumption~\ref{ass:convergence} hold, and $\rho_0^Z$ is Borel
    measurable. Then every fair play from $x_0 \in Z_\epsilon$
    consistent with $\rho_0^Z$ satisfies $\RA$. Consequently, $Z_\epsilon
        \subseteq
        \Win_0^{\mathrm f}(G(\mathcal S),\RA)$.
    }
\end{theorem}

\edit{The proof follows the finite-state argument for rank-decreasing strategies. Along a play that never reaches $T$, the rank of the current state is at least $2$, and the progress condition~\eqref{eq:rank-decreasing-input} overlaps the successor set with the interior of the next lower rank by measure more than $\epsilon$. Since $Q$ is bounded, the successor sets $F(x_i) + \W$ cannot escape (this is where boundedness is needed, see Example~\ref{ex:unbounded}), and the following lemma extracts a single basis ball inside the lower rank that is contained in infinitely many of them. This ball is then enabled infinitely often, so fairness forces the play to visit it, which decreases the rank until finally some state with rank one (and therefore in $T$) is visited.}

\begin{lemma}[Persistent ball under translations]
    \label{lem:ball-persist}
    Let $S, R \subseteq \R^n$ be bounded open sets. If $(z_k)_{k \in \N}$ is a sequence in $\R^n$ with $\leb((z_k + R) \cap S) \geq \epsilon$ for some fixed $\epsilon > 0$ and all $k$, then some basis ball $B \in \mathcal{B}$ satisfies $B \subseteq S \cap (z_k + R)$ for infinitely many $k$.
\end{lemma}

\begin{proof}
    For $\delta > 0$, define $R_\delta \defeq \{\, w \in R \mid B(w, \delta) \subseteq R \,\}$ and $S_\delta \defeq \{\, x \in S \mid B(x, \delta) \subseteq S \,\}$. Since $R$ and $S$ are bounded open sets, $\leb(R \setminus R_\delta) \to 0$ and $\leb(S \setminus S_\delta) \to 0$ as $\delta \to 0$, so we can fix $\delta > 0$ with $\leb(R \setminus R_\delta) < \epsilon/2$ and $\leb(S \setminus S_\delta) < \epsilon/2$. Then for every $k$,
    $$
        \leb\bigl((z_k + R_\delta) \cap S_\delta\bigr)
        \geq \epsilon - \leb(R \setminus R_\delta) - \leb(S \setminus S_\delta) > 0,
    $$
    so $(z_k + R_\delta) \cap S_\delta$ is non-empty. Cover the bounded set $S$ by finitely many basis balls of radius at most $\delta/2$. By the pigeonhole principle, one of these balls $B \in \mathcal{B}$ intersects $(z_k + R_\delta) \cap S_\delta$ for infinitely many $k$. Every point of $B$ lies within distance $\delta$ of that intersection point, and hence inside both $S$ and $z_k + R$. Therefore, $B \subseteq S \cap (z_k + R)$ for infinitely many $k$.
\end{proof}

\begin{proof}[Proof of Theorem~\ref{thm:sound-eps}]
    Let $\pi$ be a fair play from $x_0 \in Z_\epsilon$ consistent with $\rho_0^Z$; by Lemma~\ref{lem:safety}, it suffices to show that $\pi$ visits $T$. Suppose not. Then $x_i \in Z_\epsilon \setminus T$ and hence $r(x_i) \in \dint{2}{K}$ for all $i$. Let $m \in \dint{2}{K}$ be the smallest rank attained infinitely often along $\pi$, say at steps $(i_k)_{k \in \N}$. By~\eqref{eq:rank-decreasing-input} and $\leb(\W \setminus \interior{\W}) = 0$ (Assumption~\ref{ass:geometry}(iii)), we have $\leb\bigl((F(x_{i_k}) + \interior{\W}) \cap \interior{X_{m-1}}\bigr) > \epsilon$ for all $k$. Since $\interior{\W}$ and $\interior{X_{m-1}} \subseteq Q$ are bounded open sets, Lemma~\ref{lem:ball-persist} gives a basis ball $B \in \mathcal{B}$ with $B \subseteq \interior{X_{m-1}} \cap (F(x_{i_k}) + \interior{\W}) \subseteq \Post(x_{i_k}, u_{i_k})$ for infinitely many $k$. By fairness, $\pi$ visits $B \subseteq X_{m-1}$ infinitely often, so $r(x_i) \leq m - 1$ infinitely often, contradicting the minimality of $m$. Thus $\pi \in \RA$.
\end{proof}

\subsubsection{Soundness of $Z_0$}
\label{sec:soundness-zero}
We now establish soundness for $\epsilon = 0$.
\begin{theorem}[Soundness for $\epsilon = 0$]
    \label{thm:sound-zero}
    Let $\epsilon = 0$ and consider Assumptions~\ref{ass:geometry} and~\ref{ass:convergence} to hold. Then every fair play from $x_0 \in Z_0$ consistent with $\rho_0^Z$ satisfies $\RA$. Consequently, $Z_0 \subseteq \Win_0^{\mathrm f}(G(\mathcal S), \RA)$.
\end{theorem}

For $\epsilon = 0$, the progress condition only requires a non-empty overlap with the interior of the next lower rank, which can have arbitrarily small measure, so the argument of Section~\ref{sec:soundness-eps} fails. Instead, we use o-minimality to partition $Z_0 \setminus T$ into finitely many definable pieces on which the rank and control input are constant and $F$ is continuous. We call a basis ball whose closure lies inside one piece \emph{plain}. The strategy behaves uniformly on each plain ball. If a fair play never reaches $T$, boundedness ensures that some plain ball is enabled infinitely often. Fairness forces the play to visit this ball infinitely often, and continuity of $F$ on it then gives another infinitely-often-enabled plain ball of strictly lower rank. Since there are only finitely many ranks, this descent cannot continue, giving a contradiction. First, we establish definability and piecewise continuity on $Z_0$:
\begin{lemma}[Structure of $Z_0$]
    \label{lem:struct}
    Under Assumptions~\ref{ass:geometry} and~\ref{ass:convergence} (with $\epsilon = 0$):
    \begin{enumerate}[(i)]
        \item $Z_0$, $X_1,\dots,X_K$, and $\rho_0^Z$ are definable, and $\rho_0^Z$ is Borel.
        \item There is a finite partition $\mathcal{C}$ of $Z_0$ into definable sets such that $r$ and $\rho_0^Z$ are constant and $F$ is continuous on each $C \in \mathcal{C}$.
    \end{enumerate}
\end{lemma}
\begin{proof}
    (i) Since $\ALPre^0 = \ALPre$, for any definable sets $S, P \subseteq \R^n$,
    $\ALPre^0(S,P) = \bigcup_{u \in \U} \bigl\{\, x \in \X \;\bigm|\; \Post(x,u) \subseteq S \text{ and } \Post(x,u) \cap \interior{P} \neq \emptyset \,\bigr\}.$
    Since $\Post(x,u) = f(x,u) + \W$ and interiors of definable sets are definable, this is a first-order description over the definable objects $f(\cdot,u)$, $\W$, $S$, and $P$, so $\ALPre^0(S,P)$ is definable. Starting from $Y_0 = \X$, every iterate $Y_k$ and $X_m$ is built by finitely many applications of $\ALPre^0$ and Boolean operations (Assumption~\ref{ass:convergence}); hence all $Y_k$, $Z_0 = Y_N$, and $X_1,\dots,X_K$ are definable. Likewise, for each $u \in \U$ and $m \in \dint{2}{K}$, the set of states in $X_m \setminus X_{m-1}$ where $u$ satisfies~\eqref{eq:rank-decreasing-input} is definable. Selecting the first such $u$ in the fixed enumeration of $\U$ on $Z_0$, and a fixed input on $T$, makes $\rho_0^Z$ definable and therefore Borel.

    (ii) Partition $Z_0$ into the finitely many definable sets $R_{m,u} \defeq \{\, x \in X_m \setminus X_{m-1} \mid \rho_0^Z(x) = u \,\}$ for $m \in \dint{1}{K}$ (with $X_0 = \emptyset$) and $u \in \U$. On each $R_{m,u}$, $r$ and $\rho_0^Z$ are constant and $F = f(\cdot, u)$ is definable. Applying (P2) to each $R_{m,u}$ refines this into the desired partition $\mathcal{C}$.
\end{proof}
Using the partition $\mathcal{C}$, we call a basis ball $B_j$ \emph{plain} if $\cl{B_j} \subseteq C$ for some $C \in \mathcal{C}$. On any plain ball $B_j$, the rank $r$ and input $\rho_0^Z$ are constant (we write $r(B_j)$), and $F$ is continuous on the compact set $\cl{B_j}$. Plain balls are easy to find, in fact, every open subset of $Z_0$ contains at least one. 
\begin{lemma}[Plain balls]
    \label{lem:plain}
    For every non-empty open set $O \subseteq Z_0$, there exists a plain basis ball $B_j$ with $\cl{B_j} \subseteq O$.
\end{lemma}
\begin{proof}
    Since $O$ is open and non-empty, pick a basis ball $B$ with $\cl{B} \subseteq O$. Because $B \subseteq Z_0$ and $\mathcal{C}$ is a finite definable partition of $Z_0$, we can write $B$ as the finite union of definable sets $B = \bigcup_{C \in \mathcal{C}} (B \cap C)$. Since $B$ is a non-empty open ball, by~(P1) some $B \cap C$ also has non-empty interior $V$. Choosing a basis ball $B_j$ with $\cl{B_j} \subseteq V$ gives $\cl{B_j} \subseteq C$ (so $B_j$ is plain) and $\cl{B_j} \subseteq B \subseteq O$.
\end{proof}

Next we show that a converging sequence eventually always cover any compact subset of its limit interior:
\begin{lemma}[Limit coverage]
    \label{lem:persist}
    If $z_k \to y$ in $\R^n$ and $K \subseteq y + \interior{\W}$ is compact, then $K \subseteq z_k + \W$ for all sufficiently large $k$.
\end{lemma}
\begin{proof}
    Since $K$ is compact and lies inside the open set $y + \interior{\W}$, the distance $d$ from $K$ to the complement of $y + \interior{\W}$ is positive. Whenever $|z_k - y| < d$, every $p \in K$ satisfies $p + y - z_k \in y + \interior{\W}$, i.e., $p \in z_k + \interior{\W} \subseteq z_k + \W$.
\end{proof}

With these lemmas, we formally prove the soundness for $Z_0$. 

\begin{proof}[\edit{Proof of Theorem~\ref{thm:sound-zero}}]
    Let $\pi$ be such a fair play. By Lemma~\ref{lem:safety} it suffices to show that $\pi$ visits $T$. Suppose not. Then $x_i \in Z_0 \setminus T$ and $u_i = \rho_0^Z(x_i)$ for all $i$, so $\Post(x_i, u_i) = z_i + \W \subseteq Z_0 \subseteq Q$ where $z_i \defeq F(x_i)$. Since $Q$ is bounded (Assumption~\ref{ass:geometry}(i)) and $\W$ is non-empty, the sequence $(z_i)$ is bounded. Call a basis ball \emph{good} if it is plain and enabled infinitely often along $\pi$. Every good ball $B$ has $r(B) \in \dint{2}{K}$: otherwise $r(B) = 1$, so $B \subseteq T$ is enabled infinitely often and fairness forces $\pi$ to visit $T$, a contradiction.
    
    \emph{A good ball exists.} Take a convergent subsequence $z_{i_k} \to y$. By Lemma~\ref{lem:persist}, $y + \interior{\W} \subseteq Z_0$, so Lemma~\ref{lem:plain} gives a plain ball $B_j$ with $\cl{B_j} \subseteq y + \interior{\W}$. Applying Lemma~\ref{lem:persist} to $\cl{B_j}$ then gives $B_j \subseteq z_{i_k} + \W = \Post(x_{i_k}, u_{i_k})$ for all large $k$, so $B_j$ is good.
    
    \emph{But no good ball has minimal rank.} Let $B$ be a good ball of minimal rank $r \defeq r(B)$. By fairness, $\pi$ visits $B$ infinitely often. Because $\cl{B}$ is compact, there is a subsequence $x_{i_k} \in B$ converging to some $x^* \in \cl{B}$. Because $B$ is plain, $r$ is constant and $F$ is continuous on $\cl{B}$, so $r(x^*) = r \geq 2$ and $F(x_{i_k}) \to F(x^*)$. Because $r(x^*) = r$, the progress condition~\eqref{eq:rank-decreasing-input} ensures that $F(x^*) + \interior{\W}$ intersects with $\interior{X_{r-1}}$. Inside this non-empty open intersection, Lemma~\ref{lem:plain} finds a plain ball $B'$ (with $\cl{B'}$ still inside). Since $F(x_{i_k}) \to F(x^*)$, Lemma~\ref{lem:persist} guarantees that $B'$ is contained in $F(x_{i_k}) + \W$ for all large $k$, making $B'$ a good ball of rank at most $r - 1$ and thus contradicting the minimality of $r$.
    Hence $\pi$ visits $T$.
    
\end{proof}

\subsection{Completeness}
\label{sec:completeness}

We now show the converse inclusion for $\epsilon = 0$: every state outside $Z_0$ is losing for Player~$0$ under fair plays, i.e., against every Player~$0$ strategy the environment has a fair play that violates $\RA$. As in Section~\ref{sec:soundness}, the proof is driven by a rank, now on the complement of $Z_0$ and induced by the outer iterates, and by a Player~$1$ strategy that never increases this rank. Throughout this subsection $Y_k$, $\Phi_0$, $X_m(\cdot)$, and $N$ refer to the iterates in~\eqref{eq:outer-iterates-eps} and~\eqref{eq:inner-approximants} with $\epsilon = 0$. Since each $Y_{k+1} = \Phi_0(Y_k)$ is a fixed point of its inner operator, for every $k \in \N$
\begin{equation}
    \label{eq:Y-fixpoint}
    Y_{k+1} = T \cup \bigl( Q \cap \ALPre^{0}(Y_{k}, Y_{k+1}) \bigr).
\end{equation}

\begin{theorem}[Completeness]
    \label{thm:completeness}
    Under Assumptions~\ref{ass:geometry}(iii) and~\ref{ass:convergence},
        $\Win_{0}^{\mathrm{f}}\bigl( G(\mathcal{S}), \RA \bigr) \subseteq Z_{0}$.
    Equivalently, for every initial state $x_{0} \in \X$, if $x_{0} \notin Z_{0}$, then $x_{0} \notin \Win_{0}^{\mathrm{f}}\bigl( G(\mathcal{S}), \RA \bigr)$.
\end{theorem}

\edit{The proof mirrors the finite-state argument and constructs, for an arbitrary Player~$0$ strategy, a Player~$1$ strategy that keeps the play outside $Z_0$ and is nevertheless fair. The outer iterates induce a \emph{Player~$1$ rank} $R(x)$ on the complement of $Z_0$, the first $k$ with $x \notin Y_k$. By~\eqref{eq:Y-fixpoint}, a state in $Q$ outside $Y_k$ is outside $T$, and for every input either invariance with respect to $Y_{k-1}$ or progress with respect to $Y_k$ fails. Player~$1$ exploits whichever fails: if invariance fails, it moves outside $Y_{k-1}$ and thereby decreases the rank; otherwise progress fails, so the successor set does not intersect the interior of $Y_k$, and Player~$1$ can stay outside $Y_k$ while still choosing freely among the enabled basis balls. It uses this freedom to serve the enabled balls in a weighted round-robin fashion. Since the rank never increases, it is eventually constant, so from some step on the round-robin rule is applied indefinitely and guarantees fairness.
}

\begin{proof}[\edit{Proof of Theorem~\ref{thm:completeness}}]
\label{app:proof-completeness}

For every $x \notin Z_0$, define the \emph{Player~$1$ rank}
$$
    R(x) \defeq \min \{\, k \geq 1 \mid x \notin Y_k \,\}.
$$
Since $Z_0 = Y_N$ and $T \subseteq Y_k$ for all $k$, we have $R(x) \in \dint{1}{N}$ and $x \notin T$.

For a play $\pi$, let $N_i(j) \defeq |\{k < i \mid k \in S_j(\pi)\}|$ be the number of visits to $B_j$ before step $i$. Because $\Post(x_i, u_i)$ has non-empty interior (Assumption~\ref{ass:geometry}(iii)), it always contains at least one basis ball. Let $j^*(i)$ be the smallest $j \geq 1$ minimizing $j + N_i(j)$ over all $j$ with $B_j \subseteq \Post(x_i, u_i)$ (a minimizer exists since $j + N_i(j) \in \N \setminus \{0\}$). We say that Player~$1$ \emph{applies the round-robin rule at step $i$} if $x_{i+1} \in B_{j^*(i)}$.

\begin{lemma}[Round-robin is fair]
    \label{lem:round-robin}
    If Player~$1$ applies the round-robin rule at every step $i \geq i_0$ along a play $\pi$, then $\pi$ is fair.
\end{lemma}
\begin{proof}
    Suppose $|E_j(\pi)| = \infty$ but $|S_j(\pi)| = c < \infty$ for some $j \geq 1$. At every step $i \geq i_0$ with $i \in E_j(\pi)$, the ball $B_j$ is enabled, so the chosen index $j' \defeq j^*(i)$ satisfies
    \[
        j' + N_i(j') \leq j + N_i(j) \leq j + c,
    \]
    which implies $j' \leq j + c$ and $N_i(j') \leq j + c$. Because $x_{i+1} \in B_{j'} \subseteq \Post(x_i, u_i)$, each such step increments $N_i(j')$ by~$1$; thus each of the finitely many indices $j' \leq j + c$ can be chosen at most $j + c + 1$ times. Hence only finitely many steps $i \geq i_0$ lie in $E_j(\pi)$, a contradiction.
\end{proof}
    Fix $x_0 \notin Z_0$ and an arbitrary Player~$0$ strategy $\rho_0$. We construct a Player~$1$ strategy $\rho_1$ whose induced play $\pi$ is fair and violates $\RA$.

    \noindent\textbf{Strategy construction.}\quad
    Let $e \in \N \cup \{\infty\}$ be the first step with $x_e \notin Q$. Player~$1$ operates in two modes:

    \smallskip
    \noindent\emph{Free mode} ($i \geq e$). Player~$1$ applies the round-robin rule, picking any $x_{i+1} \in B_{j^*(i)} \subseteq \Post(x_i, u_i)$.

    \smallskip
    \noindent\emph{Pursuit mode} ($i < e$, so $x_0,\dots,x_i \in Q$). We maintain the invariant $x_i \notin Z_0$ (which holds at $i = 0$). Let $k \defeq R(x_i)$. Since $x_i \in Q \setminus Y_k$, Eq.~\eqref{eq:Y-fixpoint} gives $x_i \notin T$ and $x_i \notin \ALPre^0(Y_{k-1}, Y_k) = \ALPre(Y_{k-1}, Y_k)$. Thus, for $u_i = \rho_0(\dots)$, one of two cases holds (giving priority to (a)):
    \begin{enumerate}[(a)]
        \item $\Post(x_i, u_i) \not\subseteq Y_{k-1}$: Player~$1$ picks any $x_{i+1} \in \Post(x_i, u_i) \setminus Y_{k-1}$. Then $k \geq 2$ (as $Y_0 = \X$) and $R(x_{i+1}) \leq k - 1 < R(x_i)$.
        \item $\Post(x_i, u_i) \subseteq Y_{k-1}$ and $\Post(x_i, u_i) \cap \interior{Y_k} = \emptyset$: since $B_{j^*(i)}$ is a non-empty open subset of $\Post(x_i, u_i)$, it cannot lie inside $Y_k$ (otherwise $B_{j^*(i)} \subseteq \interior{Y_k}$). Player~$1$ picks any $x_{i+1} \in B_{j^*(i)} \setminus Y_k$, applying the round-robin rule while maintaining $R(x_{i+1}) \leq k = R(x_i)$.
    \end{enumerate}
    In both cases $x_{i+1} \in \Post(x_i, u_i)$ and $x_{i+1} \notin Z_0$ (as $Z_0 \subseteq Y_m$ for all $m$), so $\rho_1$ is legal and preserves the invariant.

    \medskip
    \noindent\textbf{The play violates $\RA$.}\quad
    For every $i < e$, we have $x_i \notin Z_0$ and hence $x_i \notin T$. If $e < \infty$, then also $x_e \notin Q \supseteq T$, so $T$ is not reached on or before step $e$, and leaving $Q$ at step $e$ prevents satisfying $\RA$ afterwards. Thus $\pi \notin \RA$.

    \medskip
    \noindent\textbf{The play is fair.}\quad
    If $e < \infty$, Player~$1$ uses free mode—and thus the round-robin rule—for all $i \geq e$. If $e = \infty$, the rank $R(x_i) \in \dint{1}{N}$ is non-increasing and hence eventually constant, say for all $i \geq i_0$; case~(a) cannot occur for $i \geq i_0$ (as it strictly decreases $R$), so Player~$1$ applies case~(b)—and thus the round-robin rule—for all $i \geq i_0$. In either case, Lemma~\ref{lem:round-robin} guarantees that $\pi$ is fair.

    Therefore $x_0 \notin \Win_0^{\mathrm{f}}(G(\mathcal{S}), \RA)$, proving $\Win_0^{\mathrm{f}}(G(\mathcal{S}), \RA) \subseteq Z_0$.
\end{proof}

\subsection{Discussion}
\label{sec:exact}
As a direct consequence of Theorem~\ref{thm:completeness} and Theorem~\ref{thm:sound-zero} we first observe that the fixed point in \eqref{eq:main-fixpoint}
indeed yields an exact characterization of the fair winning domain.

\begin{corollary}[$Z_0 = \Win_0^{\mathrm f}$]
    \label{cor:exact}
    Under Assumptions~\ref{ass:geometry} and~\ref{ass:convergence} (with $\epsilon = 0$), it holds that
    \begin{equation*}
        Z_0 = \Win_0^{\mathrm f}\bigl(G(\mathcal S), \RA\bigr) \subseteq \Win_0^{\mathrm p}\bigl(G(\mathcal S), \RA\bigr).
    \end{equation*}
\end{corollary}
We note that the inclusions ``$\supseteq$'' and ``$\subseteq$'' on the left follow from Theorem~\ref{thm:completeness} and Theorem~\ref{thm:sound-zero}, respectively, as $\rho_0^Z$ is definable, hence Borel (Lemma~\ref{lem:struct}(i)), and thus an admissible strategy. The inclusion on the right-hand side follows from Theorem~\ref{thm:hierarchy}.

\SES{Thus, the fixed point $Z_0$ is exactly the fair winning domain and the system wins from states in $Z_0$ via
the memoryless definable strategy $\rho_0^Z$ against a fair environment. Since fairness holds almost surely (Theorem~\ref{thm:prob-fairness}), the same strategy also wins
almost surely from $Z_0$. In fact, continuity of $f$ is not required for this result to follow. However, boundedness of $Q$ (Assumption~\ref{ass:geometry}(i)) is needed for soundness, while completeness does not require it. The following example shows that boundedness is indeed needed for \emph{both} soundness results, i.e., for any $\epsilon\geq 0$.}

\begin{example}[Unbounded $Q$]
    \label{ex:unbounded}
    Let $\X = \R^2$, $\U = \{u\}$, $\W = [-1,1]^2$ with $\mu$ uniform, $Q = \R^2$, $T = [-1,1]^2$, and
    \begin{equation*}
        f((x,y),u) \defeq \begin{cases} (0,0) & \text{if } y > 0, \\ (x+3,\,0) & \text{if } y \leq 0. \end{cases}
    \end{equation*}
    Then $f$ is piecewise linear, hence definable, and Assumption~\ref{ass:geometry} holds except that $Q$ is unbounded.
    \emph{$Z_0 = \R^2$.} If $y > 0$, then $\Post = [-1,1]^2$, which intersects $\interior{T}$; so $\{y > 0\} \subseteq X_2$. If $y \leq 0$, then $\Post = [x+2,x+4] \times [-1,1]$ intersects $\{y > 0\} \subseteq \interior{X_2}$; so $X_3 = \R^2 = X_4$. Hence $K = 3$, $Y_1 = Y_0 = \R^2$, and $Z_0 = \R^2$, so Assumption~\ref{ass:convergence} holds with $N = 1$.
    \emph{$(5,0) \notin \Win_0^{\mathrm f}$.} Let $\rho_1$ choose $w \equiv 0$, giving the legal play $p_i = (5+3i,\,0)$. It never enters $T$, so it violates $\RA$. It is fair: $\Post(p_i) = [5+3i+2,\,5+3i+4] \times [-1,1]$ moves off to infinity, so every bounded ball $B_j$ is enabled only finitely often. As $\U$ is a singleton, no strategy for Player~$0$ can avoid this play.
    Thus $Z_0 \not\subseteq \Win_0^{\mathrm f}$. The same computation (with $\leb(\Post \cap \interior{T}) = 4$ and $\leb(\Post \cap \{y>0\}) = 2$) gives $Z_\epsilon = \R^2$ for every $\epsilon < 2$, hence also $Z_\epsilon \not\subseteq \Win_0^{\mathrm f}$. So boundedness of $Q$ in Assumption~\ref{ass:geometry}(i) cannot be dropped from either soundness theorem.
\end{example}

\section{Experiments and Case Studies}
\label{sec:examples}

\begin{figure}[t]
    \centering
    \vspace{-20pt}
    \includegraphics[width=\columnwidth]{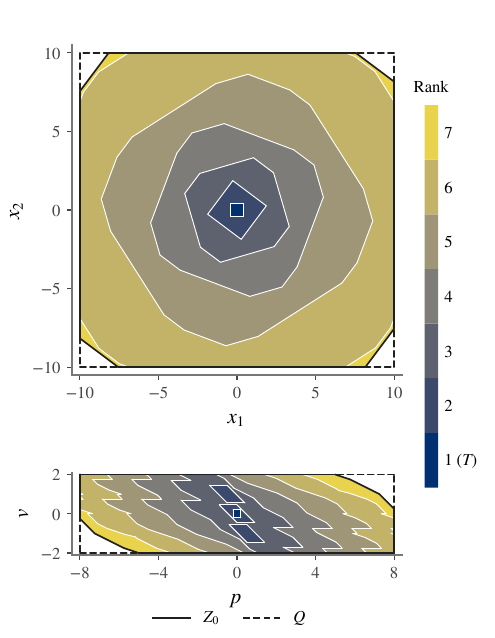}
    \caption{Fair winning domains $Z_0$ of the damped rotation  (top) and the double integrator in \eqref{eq:ex:integrator} (bottom), colored by rank. Solid line: $Z_0$; dashed line: safe set $Q$.}
    \label{fig:final-z0}
\end{figure}

We illustrate the fixed-point construction of Section~\ref{sec:synthesis} on an analytically solvable example and two numerical case studies, and we compare the result with abstraction-based synthesis. In all three systems, the disturbance set $\W$ is larger than the target $T$ in every coordinate. Hence no successor set $\Post(x,u)$ fits inside $T$, Player~$1$ can always keep the play outside $T$, and the adversarial winning domain is exactly $T$. The fair winning domain, in contrast, covers a substantial part of the safe set.

\subsection{Analytical Example}
\label{sec:ex:contractive}

Consider the linear contractive system with $\X = \R^{2}$, safe set $Q = [-2,2]^{2}$, target $T = [-0.4, 0.4]^{2}$, a single input $\U = \{u_{0}\}$ (omitted below), disturbance set $\W = [-0.5, 0.5]^{2}$, and dynamics $f(x) = 0.5x$, so that $\Post(x) = 0.5x + [-0.5, 0.5]^{2}$. We compute $Z_0$ from~\eqref{eq:inner-approximants} and~\eqref{eq:outer-iterates-eps} by hand. Note that $x \in \ALPre^{0}(Y, X)$ iff $\Post(x) \subseteq Y$ and $\interior{\Post(x)} \cap \interior{X} \neq \emptyset$. For $Y_0 = \R^2$, the inner approximants $X_m \defeq X_m(\R^2)$ are $X_1 = T$; then $X_2 = (-1.8, 1.8)^2$, since $\interior{\Post(x)} = 0.5x + (-0.5, 0.5)^2$ intersects $\interior{T} = (-0.4, 0.4)^2$ iff $|0.5 x_i| < 0.9$ for $i \in \{1,2\}$; and finally $X_3 = Q = X_4$, since $\Post(x) \subseteq [-1.5, 1.5]^2 \subset \interior{X_2}$ for every $x \in Q$. Thus $Y_1 = \Phi_0(\R^2) = Q$. Because $\Post(x) \subseteq [-1.5, 1.5]^2 \subseteq Q$ for all $x \in Q$, replacing $Y_0 = \R^2$ by $Y_1 = Q$ leaves the inner approximants unchanged, so $Y_2 = \Phi_0(Q) = Q = Y_1$. Hence Assumptions~\ref{ass:geometry} and~\ref{ass:convergence} hold with $K = 3$ and $N = 1$, and Corollary~\ref{cor:exact} gives
    $\Win_0^{\mathrm f}\bigl(G(\mathcal S), \RA\bigr) = Z_0 = Q = [-2,2]^2$,
with ranks $1$ on $T$, $2$ on $(-1.8, 1.8)^2 \setminus T$, and $3$ on $Q \setminus (-1.8, 1.8)^2$. The adversarial winning domain is only $T$, whereas the fair winning domain is the entire safe set. Since $\U$ is a singleton, the strategy $\rho_0^Z$ is trivial here and fairness alone forces progress from rank $3$ to rank $2$ and into the target.

\subsection{Numerical Case Studies}
\label{sec:ex:numerical}

For the following two systems, we compute $Z_0$ with our Python implementation, which evaluates the fixed-point iteration with exact arithmetic. Figure~\ref{fig:final-z0} shows the resulting fair winning domains together with the rank of each state. The ranks grow with the distance from the target and bound the number of fair progress steps the strategy needs to reach it.

\paragraph{Damped rotation.}
Consider the rotating contractive system with $\X = \R^{2}$, safe set $Q = [-10, 10]^{2}$, target $T = [-0.4, 0.4]^{2}$, a single input $\U = \{u_{0}\}$, disturbance set $\W = [-0.6, 0.6]^{2}$, and dynamics
\begin{equation}\label{eq:ex:rotation}
    f(x) = A x, \qquad A = \frac{3}{4} \begin{bmatrix} \cos\theta & -\sin\theta \\ \sin\theta & \cos\theta \end{bmatrix}, \qquad \theta \approx 36.87^{\circ},
\end{equation}
so that $\Post(x) = A x + [-0.6, 0.6]^{2}$. As $\W$ cannot fit inside $T$, the adversarial winning domain is $T$. The fair winning domain, shown in Figure~\ref{fig:final-z0} (top), covers a large part of the safe set. Since there is no control choice, every state in $Z_0$ is driven into $T$ by contraction and fairness alone, and the rank structure reflects the spiral motion of the dynamics.

\paragraph{Double integrator.}
Consider the discrete-time double integrator with time step $\Delta t = 1$, state $x = (p, v) \in \X = \R^{2}$, safe set $Q = [-8, 8] \times [-2, 2]$, target $T = [-0.2, 0.2]^{2}$, inputs $\U = \{-1, 0, 1\}$, disturbance set $\W = [-0.25, 0.25]^{2}$, and dynamics
\begin{equation}\label{eq:ex:integrator}
    f(x, u) = \begin{bmatrix} 1 & 1 \\ 0 & 1 \end{bmatrix} x + \begin{bmatrix} 0.5 \\ 1 \end{bmatrix} u, \qquad \Post(x, u) = f(x, u) + \W.
\end{equation}
Again $\W$ is strictly larger than $T$ in every coordinate, so the adversarial winning domain is $T$. Here the controller has a genuine choice, and the rank-decreasing strategy of Definition~\ref{def:rank-strategy} selects, in each state of $Z_0$, an acceleration that keeps all successors in $Z_0$ and makes progress towards the next lower rank. The fair winning domain is shown in Figure~\ref{fig:final-z0} (bottom).

\begin{figure}[t]
    \centering
    \includegraphics[width=\columnwidth]{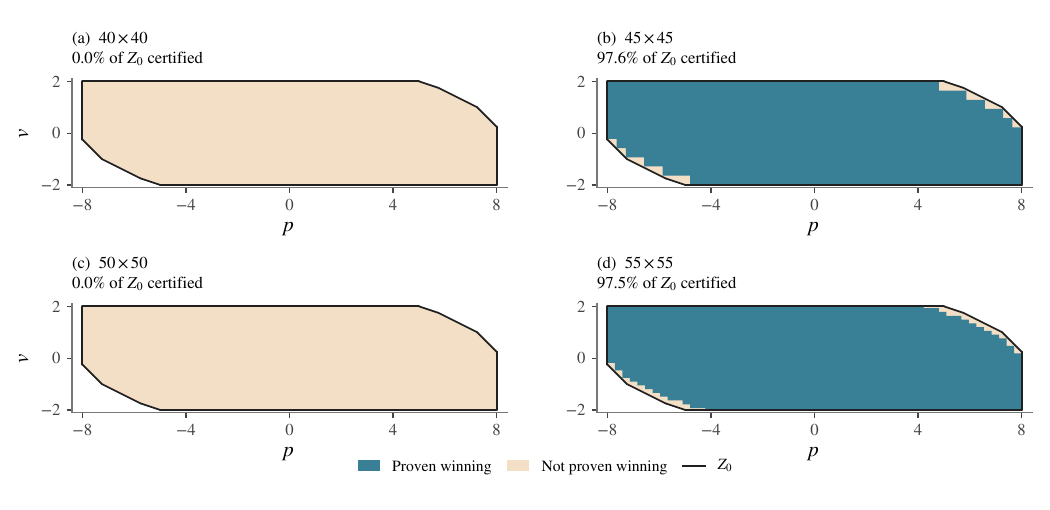}
    \caption{Double integrator: winning domains certified by the abstraction-based method of~\cite{MajumdarMallikSchmuckSoudjani2024} for four grid resolutions (blue), compared with $Z_0$ (beige).}
    \label{fig:comparison}
\end{figure}

\subsection{Comparison to Abstraction-Based Method}
\label{sec:ex:comparison}

For the double integrator, we compare $Z_0$ with the result of the abstraction-based method of~\cite{MajumdarMallikSchmuckSoudjani2024}. This method partitions the state space into a uniform grid, constructs a finite game abstraction with strong transition fairness, and solves the fair game on the abstraction, which provides an under-approximation of the fair winning domain. Figure~\ref{fig:comparison} shows the certified domains for grids of $40\times40$, $45\times45$, $50\times50$, and $55\times55$ cells. In our experiments, all certified domains lie inside $Z_0$.

The comparison exposes two problems of the detour through abstraction. The first is conservatism: even the best grid misses part of $Z_0$. The second, and more severe, is that the certified domain depends on the grid resolution in a non-monotone way. The $40\times40$ and $50\times50$ grids result in an empty winning domain, because no cell lies entirely inside the target, while the $45\times45$ and $55\times55$ grids certify $97.6\%$ and $97.5\%$ of $Z_0$, respectively. Hence, refining the grid from $45\times45$ to $50\times50$ cells reduces the certified domain from $97.6\%$ to $0\%$, while refining it further gives $97.5\%$. Our method circumvents this problem by computing $Z_0$ on the continuous domain.

\section{Conclusion}
\label{sec:conclusion}
 This paper introduced a topological notion of fairness that extends existing fairness notions to continuous-state systems. We proved that stochastic dynamical systems inherently satisfy the topological fairness assumption. We then showed that when the environment is constrained by this fairness assumption, the resulting winning domain lies between the winning domain of the adversarial setting and that of the almost-sure setting. We proposed a fixed-point algorithm that computes the winning domain and synthesizes a controller under this fairness assumption, and we proved that it recovers the fair winning domain exactly for a broad class of systems. %
 Future work will focus on two directions: (i) extending our results to richer classes of systems and specifications, and (ii) developing efficient computational methods for solving the fixed-point algorithm.

\clearpage
\bibliographystyle{ACM-Reference-Format}
\bibliography{references}
\end{document}